\documentstyle[12pt]{article}
\def\al{\alpha}
\def\be{\beta}
\def\ga{\gamma}

\def\th{\theta}

\def\la{\lambda}

\def\rh{\rho}

\def\si{\sigma}

\def\ph{\phi}
\def\vp{\varphi}
\def\ch{\chi}
\def\ps{\psi}

\def\De{\Delta}

\def\fr#1#2{{{#1} \over {#2}}}
\def\prt{\partial}
\def\ap{\al^\prime}

\def\vev#1{\langle {#1}\rangle}
\def\bra#1{\langle{#1}|}
\def\ket#1{|{#1}\rangle}

\def\half{{\textstyle{1\over 2}}}
\def\frac#1#2{{\textstyle{{#1}\over {#2}}}}

\def\lsim{\mathrel{\rlap{\lower4pt\hbox{\hskip1pt$\sim$}}
    \raise1pt\hbox{$<$}}}
\def\gsim{\mathrel{\rlap{\lower4pt\hbox{\hskip1pt$\sim$}}
    \raise1pt\hbox{$>$}}}
\def\sqr#1#2{{\vcenter{\vbox{\hrule height.#2pt
         \hbox{\vrule width.#2pt height#1pt \kern#1pt
         \vrule width.#2pt}
         \hrule height.#2pt}}}}

\def\cC{{\cal C}}
\def\cD{{\cal D}}

\def\cH{{\cal H}}
\def\cR{{\cal R}}
\def\va{\vev{A}}
\def\vb{\vev{B}}

\def\vx{\vev{X}}
\def\vp{\vev{P}}
\def\oc{\widehat{\cos\ph}}
\def\os{\widehat{\sin\ph}}

\newcommand{\beq}{\begin{equation}}
\newcommand{\eeq}{\end{equation}}
\newcommand{\bea}{\begin{eqnarray}}
\newcommand{\eea}{\end{eqnarray}}
\newcommand{\rf}[1]{(\ref{#1})}
\newcommand{\eq}[1]{Eq.\ \rf{#1}}

\renewenvironment{thebibliography}[1]
 { \rm
   \begin{list}{\arabic{enumi}.}
    {\usecounter{enumi} \setlength{\parsep}{0pt}
     \setlength{\itemsep}{3pt} \settowidth{\labelwidth}{#1.}
     \sloppy
    }}{\end{list}}

\begin{document}
\titlepage

\begin{flushright}
{IUHET 292\\}
\end{flushright}
\vglue 1cm

\begin{center}
{{\bf MINIMUM-UNCERTAINTY ANGULAR WAVE PACKETS\\}
{\bf AND QUANTIZED MEAN VALUES\\}
\vglue 1.0cm
{V. Alan Kosteleck\'y and Bogdan Tudose\\}
\bigskip
{\it Physics Department\\}
\medskip
{\it Indiana University\\}
\medskip
{\it Bloomington, IN 47405, U.S.A.\\}
\vglue 0.3cm

\vglue 0.8cm
}
\vglue 0.3cm

\end{center}

{\rightskip=3pc\leftskip=3pc\noindent
Uncertainty relations between a bounded coordinate operator
and a conjugate momentum operator
frequently appear in quantum mechanics.
We prove that physically reasonable
minimum-uncertainty solutions to
such relations have quantized expectation values
of the conjugate momentum.
This implies,
for example,
that the mean angular momentum
is quantized for any minimum-uncertainty state
obtained from any uncertainty relation
involving the angular-momentum operator
and a conjugate coordinate.
Experiments specifically seeking to create
minimum-uncertainty states localized in angular coordinates
therefore must produce packets with integer angular momentum.

}

\vskip 1truein
\centerline{\it Accepted for publication in Physical Review A}

\vfill
\newpage

\baselineskip=20pt

The use of angular-coordinate and phase operators
in quantum mechanics requires more care
than perhaps might be expected.
Consider the relatively simple example of a particle
moving on a circle of unit radius.
Classically, a point particle is
necessarily located at a single value of
the periodic angular coordinate $\ph_p \in (-\pi,\pi]$.
The corresponding quantum wave function,
however,
is an object extended around the circle
and so can be directly affected by the nontrivial topology.
A simple consequence is that,
in contrast to the continuous spectrum
of the linear momentum of a free particle
on a (topologically trivial) line,
the angular momentum $L\equiv -i \prt_\ph$
(in units with $\hbar = 1$)
on the circle has a point spectrum
imposed by the dual requirements of continuity and periodicity
of the eigenstates.
A more subtle consequence is difficulty
in defining an angular-coordinate operator.
The direct use at the quantum level
of the classical coordinate $\ph_p$
causes trouble at $\ph_p = \pi$,
where any derivatives of $\ph_p$ acquire
delta-function contributions from the discontinuity.
For example,
the Dirac-type
\cite{d,h}
commutator relation $[L,\ph] = -i$
is problematic for $\ph\equiv \ph_p$.
The attempt to circumvent this by using
instead a continuous variable
$\ph \equiv \ph_c \in (-\infty, \infty)$
is also unsatisfactory
because single-valuedness restricts the Hilbert space
to the subspace of $2\pi$-periodic functions,
which \it inter alia \rm
excludes the coordinate $\ph_c$ as an observable.

Many of the difficulties arising from the use of
$\ph_p$ or $\ph_c$
can be sidestepped by selecting instead
angular coordinates that are \it both \rm
periodic and continuous.
However,
a single such quantity cannot uniquely specify
a point on the circle because periodicity implies extrema,
which excludes a one-to-one correspondence and hence
is incompatible with uniqueness.
A relatively simple choice
\cite{l}
is to adopt \it two \rm angular-coordinate operators
$\oc$ and $\os$,
defined to satisfy the commutation relations
\beq
[L, \oc ] = i ~\os
\quad , \qquad
[L, \os ] = -i ~\oc
\quad .
\label{commrel}
\eeq
Classically,
this would correspond to the identification
$(x,y) \to (\cos\ph, \sin\ph)$
on the unit circle.

Even with periodic and continuous coordinates,
some difficulties may remain.
In the Hilbert space of
square-integrable functions on the circle,
where the angular-momentum operator
is unbounded below and above,
the action of the operators $\oc$ and $\os$ on a state
can be defined by multiplication by $\cos\ph$ and $\sin\ph$,
respectively.
This implies the commutator $[\oc,\os] = 0$.
However,
in a more general context
where the operator conjugate to the coordinates
is bounded below or above,
the introduction of suitable coordinate operators
involves further subtleties.
An example is the harmonic-oscillator number operator $N$,
for which the coordinate operators $\oc$ and $\os$
defined by analogues of \eq{commrel}
do not commute
\cite{sg,cn,rj,lhw,p,ll,n,bp,nfm}.
Moreover,
the use of cosine and sine
operators may in general be inadequate to treat all
physically interesting quantities
\cite{m}.

Uncertainty relations involving the angular momentum
are affected by the choice of angular-coordinate operators.
Define for each operator $X$ and state $\ket{\ch}$
the uncertainty
\beq
\De X \equiv \left[ \vev{X^2} - \vev{X}^2 \right]^{1/2}
\quad .
\label{unc}\eeq
The Robertson-type
\cite{r}
uncertainty relation,
\beq
\De L ~\De \ph_p \ge \half
\quad ,
\label{wrong}
\eeq
is incorrect for several reasons.
For example,
since $\De \ph_p$ is bounded within a maximum,
there is a physically acceptable limit
with sufficiently small $\De L$
that produces a violation of the inequality.
An alternative for $\ph \equiv\ph_p$ involves
modifying the definition of $\De\ph_p$
or taking into account the appearance
of delta-function contributions at $\ph_p = \pi$.
This yields a more complicated form
of the uncertainty relation
\cite{j,bmv,em},
\beq
\De L ~\fr{\De \ph_p}{1 - 3(\De\ph_p)^2/\pi^2} \ge \half f(\De\ph_p)
\quad ,
\label{jl}
\eeq
where $f(\De\ph_p)$ varies from $f = 1$
at $\De\ph_p = 0$
to $f \simeq 4.375$
at $\De\ph_p = \pi/\sqrt{3}$.
If instead one chooses the angular coordinates $\oc$ and $\os$,
then the commutators \rf{commrel} result in yet another set
of angular-momentum uncertainty relations,
\beq
\De L ~\De \oc \ge \half |\vev{\os}|
\quad ,
\qquad
\De L ~\De \os \ge \half |\vev{\oc}|
\quad .
\label{ursc}
\eeq

In the present work,
we focus in particular on
minimum-uncertainty solutions to uncertainty relations
involving angular or phase coordinates
\cite{fn}.
In topologically trivial situations
such as the free particle or the harmonic oscillator,
minimum-uncertainty solutions to the
uncertainty relations for the position and momentum operators
$X$ and $P$ exist in the Hilbert space
for arbitrary finite expectation values
of $\vx$ and $\vp$.
Naively,
one might expect that minimum-uncertainty packets
for angular coordinates and their conjugate momentum
also have continuously adjustable mean values.
However,
this intuition is incorrect.
Although the detailed form
of the minimum-uncertainty solutions depends
on the particular angular-coordinate uncertainty relations
involved,
it turns out that minimum-uncertainty packets
generically exist only for
\it quantized \rm expectation values
of the associated conjugate momentum.

In what follows,
we provide a detailed proof of this result
and discuss some consequences.
We begin with a discussion of limitations
on the uncertainty of a general self-adjoint operator,
and then turn our attention to the stronger constraint
that follows for certain operators
from the requirement of minimum uncertainty
imposed via an uncertainty relation.
Finally,
we discuss some examples
and extensions of the results.

The analysis uses a number of basic results
in the theory of linear operators on Hilbert spaces
\cite{jw}.
The following summarizes some of our notation.
Let $A$ be a self-adjoint operator in a Hilbert space
with domain $\cD (A)$ and let $z$ be a complex number
that is \it not \rm an eigenvalue of $A$.
Then,
the associated resolvent operator
$R(z,A) \equiv (A - z)^{-1}$
is well defined.
The resolvent set $\rh$
is comprised of those values of $z$
for which $R(z,A)$ is bounded.
The spectrum of $A$
is $\si\equiv \cC/\rh$.
It contains two components:
the point spectrum $\si_p$
consisting of the eigenvalues of $A$;
and the remainder,
which we call the continuum
and denote by $\si_c$.
We refer to the spectrum as discrete
in the special case where it
consists entirely of isolated eigenvalues
with finite multiplicities.
Also,
by definition the extended discrete spectrum of
an unbounded operator
differs from the spectrum by the addition
of the point at infinity.

Consider a self-adjoint operator $A$ on a
Hilbert space $\cH$.
For an element $\ps\in \cD (A^2)$,
the uncertainty $\De A_\ps$ defined
by \eq{unc} is the norm
$||(A - \vev{A})\ket{\ps}||$.
First,
we examine the possibility of varying
to zero the uncertainty while keeping
fixed the expectation.
We are therefore interested in
a sequence $\{\ket{n}\}$
of unit-normalized states convergent
in the Hilbert space such that
$\vev{A} \equiv \bra{n}A\ket{n} \equiv \al \in \cR$
is constant and such that
lim$_{n \to \infty}\De A_n = 0$.

Now,
$\al$ is
in the point spectrum $\si_p$,
in the continuum $\si_c$,
or in the resolvent $\rh$.
If $\al$ is in the point spectrum,
there must exist a unit-normalized state
$\ket{\ps}\in \cD (A)$
such that $A\ket{\ps} =\al\ket{\ps}$.
It is then in general possible to find
convergent sequences of the desired type.
In particular,
the constant sequence $\{\ket{n}=\ket{\ps}~ \forall ~n \}$
satisfies the requirements.
Nonconstant convergent sequences may also exist.
If, however, $\al$ is in the continuum $\si_c$
then there may exist sequences
with uncertainty approaching zero
but they cannot converge.
This follows from a theorem
for self-adjoint operators
on Hilbert spaces
\cite{jw}
stating that if there exists a convergent sequence
of unit-normalized states such that
$(A - \al)\ket{n}\to 0$ as $n\to \infty$,
then $\al\in\si_p$.
Moreover,
if instead $\al$ lies in the resolvent $\rh$,
the desired sequences cannot exist.
This follows from another theorem for self-adjoint operators
\cite{jw}
stating that for unit-normalized states
in the domain of $A$ and for $\al \in \rh$ the norm
$||(A - \al)\ket{n}||$ is greater than a positive constant $c$.
One way of seeing this last result is that if
such a sequence did exist
then in the limit $\De A \to 0$
for which $c \to 0$
the resolvent operator $R(\al,A)$
could not be bounded,
contradicting the assumption
$\al\in\rh$.

The above results already establish,
independently of any uncertainty relation,
that the uncertainty of a self-adjoint operator
cannot be dialed to zero
while maintaining constant expectation
unless the expectation value is an eigenvalue.
{}From the physical viewpoint,
however,
there are two reasons
why this is less restrictive than it might appear.
First,
the result does not preclude the possibility
that for a given situation
the uncertainty could be dialed close to zero,
i.e., the constant $c$ might be very small.
Second,
the result assumes convergence of the
sequence in the Hilbert space,
which may not be the case in all situations
of physical interest.
For example,
the position operator on the line
has no point spectrum
and therefore by the above argument
no constant-expectation packet can be
constructed with uncertainty that can be varied to
reach zero.
Thus,
the above argument
excludes a freely evolving gaussian packet
because the limiting state is a delta-function,
which is not a state in the Hilbert space.

We next show that for certain operators
the result can be substantially strengthened
if we take into account minimum-uncertainty constraints
arising from an uncertainty relation.
We are interested in particular
in the situation of minimum-uncertainty wave packets
for which the uncertainty relation
connects a bounded operator $B$
with an operator $A$ having a discrete spectrum.
Operators of these types frequently occur in physics.
For example,
$B$ could represent the coordinate operator on
a compact manifold without boundary,
such as the $n$-sphere,
with $A$ being the associated operator for
the (angular) momentum.

Consider two self-adjoint operators $A$ and $B$,
defined in a Hilbert space $\cH$ and satisfying
\beq
[A,B] = iC
\quad
\eeq
on a subspace of $\cH$.
We assume $A$ has a discrete spectrum
and $B$ is bounded.
The uncertainty relation obeyed by these operators
for any given state is
\cite{mess}
\beq
\De A ~ \De B \ge \half |\vev{C}|
\quad .
\label{ur}
\eeq

We seek minimum-uncertainty solutions $\ket{\ps}$
of \eq{ur},
defined as normalized solutions of the limiting equality.
In general,
such solutions are called squeezed states
\cite{fn2}.
They are in one-to-one correspondence
with solutions $\ket{\ps}$ of the equation
\cite{mess}
\beq
\left( A - \al \right) \ket{\ps} =
i S \left( B - \be \right) \ket{\ps}
\quad ,
\label{s}
\eeq
where $\al = \va$, $\be = \vb$,
and
\beq
S\equiv \De A/\De B = |\vev{C}|/2 (\De B)^2
\quad
\label{sq}
\eeq
is a real constant called the
squeezing.
The reality of $\al$, $\be$, and $S$
follows from the assumption that $A$, $B$, and $C$ are
self-adjoint (observable).

Suppose $\al$ is in the resolvent set $\rh$ of $A$.
Then,
\eq{s} can be expressed in terms of
the resolvent operator $R(\al,A)$ as
\beq
\left[ i S R(\al,A) (B - \be) - 1 \right]\ket{\ps} = 0
\quad .
\eeq
This is an eigenvalue equation for the operator
$X\equiv i S R(\al,A) (B-\be)$.
Since $A$ has discrete spectrum and
since the spectrum $\si(R)$ of the resolvent operator
$R(\al,A)$ can be obtained
\cite{tk}
from the extended spectrum $\tilde \si(A)$
of $A$ via the transformation
$f(\la) = (\la - \al)^{-1}$ for $\la \in \tilde\si (A)$,
it follows that
$\si(R)$ is bounded and
consists of isolated eigenvalues of finite multiplicity
possibly accumulating only at zero.
The resolvent operator $R(\al,A)$ is therefore compact.
Since by assumption $B$ is bounded,
it follows that $X$ itself is also compact.
Thus,
the spectrum of $X$ is also
bounded and consists of isolated eigenvalues
with finite multiplicity that can accumulate only at zero.
Since one must be an eigenvalue of $X$,
this means that there can be at most a finite number of
squeezed states with fixed $\al$ and $\be$
for which $S$ lies below any given value.
In fact,
not only are such solutions finite in number
but they are also generically rare:
they are in one-to-one correspondence with
the eigenvalues of $R(\al,A) (B-\be)$
lying on the imaginary axis of the complex plane.

For $\va\in\rh$,
the above argument excludes the existence
of squeezed states allowing the dialing
of $S$ and hence of $\De A$ to zero.
In fact,
the same argument shows
there is no possibility in \it any \rm
interval of dialing the
minimum-uncertainty values $\De A$ at constant $\va$,
unless $\va$ is in the spectrum of $A$.
For an operator that has a spectrum without a continuum,
it follows that any physically reasonable
minimum-uncertainty wave packet
must have quantized mean value.
In particular,
the angular momentum of a minimum-uncertainty
wave packet on the circle must be integer valued.

With the main proof completed,
we can illustrate some of the ideas
via an explicit example.
Consider the special case of circular squeezed states
(CSS)
used in ref.\ \cite{bkt} for the construction of
elliptical squeezed states
following a classical keplerian orbit
in a planar Rydberg atom.
The CSS are minimum-uncertainty states
in the Hilbert space $\cH = L^2(S^1)$ on the circle,
determined from the uncertainty relations \rf{ur}
with the identifications
$A\equiv L = -i\prt_\ph$,
$B\equiv \os$,
and
$C\equiv \oc$,
where $\os$ and $\oc$ are defined via multiplication
by $\sin\ph$ and $\cos\ph$, respectively.
To obtain a packet centered about the point
with $x=1$ on the unit circle,
we impose $\vev{\os}=0$, $\vev{\oc}>0$.
The solution of \eq{s} is
\beq
\ket{S} =
\sqrt{\fr 1 {2 \pi I_0 (2S)}}
\exp(S \cos \ph + i \vev{L} \ph)
\quad ,
\label{css}
\eeq
where $S$ is the squeezing.
In the limit $S\to 0$,
\beq
\ket{S} \to \ket{S_0} =
\sqrt{\fr 1 {2\pi}}
\exp(i \vev{L} \ph)
\quad ,
\label{lim}
\eeq
which is an angular-momentum eigenstate
with eigenvalue $\vev{L}$.
According to our result,
$\vev{L}$ must be integer
for the solutions $\ket{S}$ to exist.
In the context of the limit in \eq{lim},
this agrees with the usual requirement of
quantized eigenvalues for eigenstates of $L$.
It can also be seen directly from \eq{css}
to be necessary since $\ket{S}$
is required to be periodic in $\ph$ with period $2\pi$
to remain consistently defined
\cite{cn2}.

As another example,
consider the number operator $N$ for the harmonic oscillator.
This operator has a discrete spectrum.
Our analysis therefore shows that
the expectation $\vev{N}$ is integer valued
for any state minimizing an uncertainty relation involving $N$
and a (bounded) phase operator.

The stronger constraints we have obtained above
make use of the uncertainty relation \eq{ur}.
For the special case where $A$ is the angular-momentum
operator on the circle,
the alternative form \rf{jl}
of the uncertainty relation might be considered instead.
In this equation,
$\De L$ may be interpreted as usual,
but the definition of $\De\ph_p$
involves determining the minimum in a real parameter $\ga$
of a functional of $\ph_p$:
\beq
(\De\ph_p)^2 \equiv
{\rm min}_\ga \int_{-\pi}^{\pi} d\ph_p~
\ps^*(\ph_p +\ga) ~\ph_p^2 ~\ps(\ph_p + \ga)
\quad .
\eeq
Next,
we show that the requirement of discrete $\vev{L}$
extends to this case also.

Let a general wave function on the circle be written as
\beq
\ps(\ph_p) = r(\ph_p) \exp [i \th (\ph_p)]
\quad ,
\eeq
with $r(\ph_p)$ and $\th(\ph_p)$ both real functions,
not necessarily positive.
Suppose $r(\ph_p)$ is kept fixed,
which also fixes $\De\ph_p$.
Then, the minimum value of $\De L$
can be obtained by varying with respect to $\th(\ph_p)$.
A short calculation shows that the minimum is attained
if $\th(\ph_p)$ is linear in $\ph_p$.
The requirement of periodicity on $\ps$ then restricts
the possible choices to $\th (\ph_p) = m \ph_p + k$,
where $m$ is integer or half-integer and $k$ is a constant.
For these choices,
it can also be shown that $\vev{L} = m$.
This means that any minimum-uncertainty packet
based on \rf{jl}
must also have a discrete value of $\vev{L}$.

The case of integer $m$ has been pursued in detail in
ref.\ \cite{em},
where solutions are shown to exist
and $f(\De\ph_p)$ is numerically computed.
In this case,
the value of $\De L$
can be continuously dialed to zero
for solutions with fixed integer $\vev{L}$.
Whether or not solutions exist for half-integer $m$,
they could not have a value of $\De L$ less than
the limiting value.
In this case, the limiting value turns out to be $\De L = 1/2$.

The suggestion has also been made of combining the
idea of well-defined angular coordinates
with the idea of more complicated uncertainty relations
\cite{ah}.
The uncertainty in the conjugate coordinate
to the angular-momentum operator
is defined as
\beq
\De\ph \equiv
\sqrt{\fr
{(\De \oc)^2 + (\De \os)^2}
{\vev{\oc}^2 + \vev{\os}^2} }
\quad ,
\eeq
which ranges from zero to infinity.
The associated uncertainty relation is
\beq
\De L \De \ph > \half
\quad ,
\label{x}
\eeq
a relation that follows directly
\cite{cn}
from the two uncertainty relations
\rf{ursc}.
In Eq.\ \rf{x},
the value $\half$ cannot be attained.
This is a reflection of the impossibility of
simultaneously solving for the two equalities
\rf{ursc},
which in turn is a reflection of rotational invariance
\cite{bkt}.
Following the construction of the CSS,
we can seek a minimum-uncertainty solution
centered about $x=1$ by setting $\vev{\os}= 0$.
Equation \rf{x} then reduces to a weakened version of
the second inequality
in Eq.\ \rf{ursc},
which itself must still hold.
So the corresponding minimum-uncertainty solution
is again a CSS,
which has quantized values of $\vev{L}$.

The general result that any minimum-uncertainty
state has discrete angular-momentum expectation
holds for any macroscopic body viewed
as the limit of a large quantum system.
Taken to the macroscopic limit
this means,
for example,
that any macroscopic object
either has an integer-valued angular momentum
or is not in a state of minimum angular uncertainty.

The issue of the physically correct form
of the uncertainty relations
and its implications is of more than
theoretical interest.
In recent years,
striking results have been found in the
behavior of minimum-uncertainty packets.
For example,
minimum-uncertainty radial electron wave packets in Rydberg atoms
have been shown to evolve through a complex sequence of
revivals, fractional revivals, and superrevivals
\cite{ps,az,ap,nau,bk}.
Specific attention has been given recently
to creating in the laboratory packets
with minimum uncertainty in an angular coordinate.
For example,
experimental efforts are presently underway
to produce such packets
orbiting the nucleus of a Rydberg atom
\cite{ys,gns}.
Localized packets with noninteger
angular-momentum expectation
can certainly be created,
for example,
by superposing angular-momentum eigenstates with
a gaussian weighting of coefficients
with noninteger mean value.
However,
our analysis shows
that any such packets cannot have
minimum angular uncertainty.
Experiments performed to study the classical limit
of quantum mechanics via the behavior of
minimum-uncertainty angular packets
must of necessity involve states with integer angular momentum.

We thank R. Bluhm for discussion and a critical reading
of the manuscript.

\baselineskip=19pt

\def\ajp #1 #2 #3 {Am.\ J.\ Phys.\ {\bf #1}, #3 (19#2)}
\def\ant #1 #2 #3 {At. Dat. Nucl. Dat. Tables {\bf #1}, #3 (19#2)}
\def\ap #1 #2 #3 {Ann.\ Phys.\ {\bf #1}, #3 (19#2)}
\def\fp #1 #2 #3 {Fortschr.\ Phys.\ {\bf #1}, #3 (19#2)}
\def\jmp #1 #2 #3 {J.\ Math.\ Phys.\ {\bf #1}, #3 (19#2)}
\def\jpa #1 #2 #3 {J.\ Phys.\ A {\bf #1}, #3 (19#2)}
\def\jpb #1 #2 #3 {J.\ Phys.\ B {\bf #1}, #3 (19#2)}
\def\jpsj #1 #2 #3 {J.\ Phys.\ Soc.\ Japan {\bf #1}, #3 (19#2)}
\def\mpl #1 #2 #3 {Mod.\ Phys.\ Lett.\ A {\bf #1}, #3 (19#2)}
\def\nat #1 #2 #3 {Nature {\bf #1}, #3 (19#2)}
\def\nc #1 #2 #3 {Nuov.\ Cim.\ {\bf A#1}, #3 (19#2)}
\def\ncl #1 #2 #3 {Lett.\ Nuov.\ Cim. {\bf #1}, #3 (19#2)}
\def\nim #1 #2 #3 {Nucl.\ Instr.\ Meth.\ {\bf B#1}, #3 (19#2)}
\def\npb #1 #2 #3 {Nucl.\ Phys.\ {\bf B#1}, #3 (19#2)}
\def\p #1 #2 #3 {Physics {\bf #1}, #3 (19#2)}
\def\pl #1 #2 #3 {Phys.\ Lett. {\bf #1}, #3 (19#2)}
\def\pla #1 #2 #3 {Phys.\ Lett.\ A {\bf #1}, #3 (19#2)}
\def\plb #1 #2 #3 {Phys.\ Lett.\ B {\bf #1}, #3 (19#2)}
\def\pr #1 #2 #3 {Phys.\ Rev.\ {\bf #1}, #3 (19#2)}
\def\pra #1 #2 #3 {Phys.\ Rev.\ A {\bf #1}, #3 (19#2)}
\def\prd #1 #2 #3 {Phys.\ Rev.\ D {\bf #1}, #3 (19#2)}
\def\prep #1 #2 #3 {Phys.\ Rep.\ {\bf #1}, #3 (19#2)}
\def\prl #1 #2 #3 {Phys.\ Rev.\ Lett.\ {\bf #1}, #3 (19#2)}
\def\prs #1 #2 #3 {Proc.\ Roy.\ Soc.\ (Lon.) {\bf A#1}, #3 (19#2)}
\def\ps #1 #2 #3 {Physica Scr.\ {\bf #1}, #3 (19#2)}
\def\ptp #1 #2 #3 {Prog.\ Theor.\ Phys.\ {\bf #1}, #3 (19#2)}
\def\rmp #1 #2 #3 {Rev.\ Mod.\ Phys. {\bf #1}, #3 (19#2)}


\begin{thebibliography}{xx}
\bibitem{fn2}
In certain situations
(e.g., the position-momentum uncertainty relation
for the harmonic oscillator)
the solutions are coherent states,
defined as preserving minimum uncertainty for all time.
\bibitem{cn2}
A solution of the form \rf{css} is presented in ref.\ \cite{cn}
with a footnote suggesting arbitrary $\vev{L}$ is
acceptable provided attention is restricted to
a single Riemann sheet.
However, this implies a discontinuous wavefunction
at the origin unless $\vev{L}$ is integer valued.
\end{thebibliography}
\end{document}